\documentclass{article}
\usepackage{DaringFed_20250508}

\usepackage{times}
\usepackage{soul}
\usepackage{url}
\usepackage[hidelinks]{hyperref}
\usepackage[utf8]{inputenc}
\usepackage[small]{caption}
\usepackage{graphicx}
\usepackage{amsmath}
\usepackage{amsthm}
\usepackage{booktabs}
\usepackage[switch]{lineno}

\usepackage{amssymb} 
\usepackage{multirow}

\newtheorem{myDef}{Definition}
\newtheorem{myThe}{Theorem}
\newtheorem{myLem}{Lemma}

\newtheorem{myAss}{Assumption}

\usepackage[linesnumbered,ruled,vlined]{algorithm2e}
	\usepackage{algpseudocode}
	
	\SetKwRepeat{doWhile}{do}{while}

\title{DaringFed: A Dynamic Bayesian Persuasion Pricing for Online Federated Learning under Two-sided Incomplete Information}

\author{
Yun Xin$^1$
\and
Jianfeng Lu$^{1,2}$\footnotemark[1]\and
Shuqin Cao$^3$\and
Gang Li$^4$\and
Haozhao Wang$^5$\And
Guanghui Wen$^{6}$\footnotemark[1]\\
\affiliations
$^1$School of Computer Science and Technology, Wuhan University of Science and Technology\\
$^2$Key Laboratory of Social Computing and Cognitive Intelligence (Dalian University of Technology), Ministry of Education, China\\
$^3$Hubei Province Key Laboratory of Intelligent Information Processing and Real-time Industrial System, Wuhan University of Science and Technology, China\\
$^4$College of Computer Science, Inner Mongolia University, China\\
$^5$School of Computer Science and Technology, Huazhong University of Science and Technology, China\\
$^6$School of Automation, Southeast University, China\\
\emails
yunxin.wust@gmail.com, 
\{lujianfeng, shuqincao\}@wust.edu.cn, 
gli@imu.edu.cn, 
hz\_wang@hust.edu.cn, 
ghwen@seu.edu.cn
}

\begin{document}

\maketitle

\begin{abstract}
	
	Online Federated Learning (OFL) is a real-time learning paradigm that sequentially executes parameter aggregation immediately for each random arriving client. To motivate clients to participate in OFL, it is crucial to offer appropriate incentives to offset the training resource consumption. However, the design of incentive mechanisms in OFL is constrained by the dynamic variability of Two-sided Incomplete Information (TII) concerning resources, where the server is unaware of the clients’ dynamically changing computational resources, while clients lack knowledge of the real-time communication resources allocated by the server. To incentivize clients to participate in training by offering dynamic rewards to each arriving client, we design a novel \underline{D}ynamic B\underline{a}yesian pe\underline{r}suasion pric\underline{ing} for online \underline{Fed}erated learning (DaringFed) under TII. Specifically, we begin by formulating the interaction between the server and clients as a dynamic signaling and pricing allocation problem within a Bayesian persuasion game, and then demonstrate the existence of a unique Bayesian persuasion Nash equilibrium. By deriving the optimal design of DaringFed under one-sided incomplete information, we further analyze the approximate optimal design of DaringFed with a specific bound under TII. Finally, extensive evaluation conducted on real datasets demonstrate that DaringFed optimizes accuracy and converges speed by $16.99\%$, while experiments with synthetic datasets validate the convergence of estimate unknown values and the effectiveness of DaringFed in improving the server’s utility by up to $12.6\%$.
	
\end{abstract}

\footnotetext[1]{Corresponding Author.}

\section{Introduction}

Online Federated Learning (OFL) extends Federated Learning (FL) by allowing numerous clients to collaboratively train a global model in a sequential manner, enabling real-time model updates as a random client arrives \cite{b6}. The efficient leverage of computation and communication resources while ensuring data privacy, OFL is emerging as a crucial research direction \cite{b5,b7}. Since training consumes clients' computation and communication resources, establishing an efficient compensation to encourage clients collaboration is a key issue for the success of OFL \cite{b34}. 

To address the challenges arising from dynamically varying resources and real-time reward allocation requirements in the context of incomplete information, it is crucial to focus on client incentive issues in OFL. Some literatures optimistically assume that the server is aware of all clients’ local computation resources \cite{b2,b3,b4}, or relaxes this assumption by assuming that the server is only aware of their distribution \cite{b9}. Other literatures hypothesize that the amount of communication resources is publicly available to clients \cite{b10,b11,b12}. In summary, existing literatures commonly make assumptions to mitigate the dynamic nature and incompleteness in OFL, which are difficult to hold in practical scenarios and hinder its sustainable development.

Given the factors mentioned, researching an efficient incentive mechanism in OFL is both urgent and challenging, especially under dynamically varying Two-sided Incomplete Information (TII), where both the server and clients possess constantly changing incomplete information about each other. First, \textit{clients lack knowledge of the dynamically varying available communication resources} \cite{b8,b9,b13}. Due to dynamic network conditions and the server’s privacy concerns, clients have only statistical information about the communication resources allocated by the server. In this context, the server strategically disclose partial information to each arriving client to influence their participation decisions and achieve favorable outcomes. However, ensuring clients trust the disclosed information and align their actions with the server’s expectations remains challenging. Second, \textit{the real-time computation resources owned by the clients are unknown to the server} \cite{b14,b15,b16}. The distribution of clients’ computation resources is difficult for the server to estimate and is influenced by dynamic changes due to factors such as varying geographical locations and fluctuating device battery levels over time. Without knowledge of this distribution, estimating it based on the constantly updated historical information as accurately as possible remains a challenge.

To overcome the abovementioned challenges, we design a novel \underline{D}ynamic B\underline{a}yesian pe\underline{r}suasion pric\underline{ing} for online \underline{Fed}erated learning under TII, named DaringFed, which aims to persuade clients to participate in training by strategically disclose its private available communication resources and offering dynamic rewards to each arriving client. Specifically, in the design of DaringFed, the Bayesian persuasion signal rule allows the server to modify the posterior distribution of communication resources on the client side, thereby maximizing the server’s utility by persuading clients to make decisions that are favorable to the server. The dynamic pricing rule transforms the problem of lacking distribution knowledge of clients’ local computation resources into a multi-armed bandit problem, and applies a confidence bound policy to estimate this distribution, thus dynamically incentivizing and filtering client participation in OFL. We summarize our main contributions as follows:
\begin{itemize}
	\item Theoretically, to maximize the server’s utility in OFL under TII, we incorporate Bayesian persuasion and dynamic pricing into the mechanism design. Bayesian persuasion influences clients’ beliefs about communication resources and modify their participation decisions by sending signals from the server. Meanwhile, dynamic pricing adapts rewards based on historical client decision-making, without prior knowledge of the clients.
	
	\item Methodologically, we start by modeling the interaction between the server and clients as a Bayesian persuasion game within the TII context, and subsequently prove the existence of a unique Bayesian persuasion Nash equilibrium. Following this, we design an approximate optimal DaringFed mechanism to approximate the maximization of the server’s utility. Finally, we prove that there exists a specific bound on the difference between the approximate optimal solution and the optimal one.
	
	\item Experimentally, extensive experiments conducted on real datasets show that DaringFed optimizes accuracy and converges speed by $16.99\%$. On synthetic datasets, we demonstrate that the approximate solution in DaringFed can converge, and the server’s utility can be improved by up to $12.6\%$ using the proposed mechanism.
\end{itemize}

\section{Related Work}

\textbf{Incentive Mechanism for FL.} Since clients are resource-constrained and unwilling to participate in FL training, it is necessary to motivate and attract high-quality clients through a well-designed incentive mechanism \cite{b28}. For example, \citeauthor{b4} [\citeyear{b28}] proposed an asynchronous FL scheme that sequentially selects clients and estimates the reward value by integrating cooperative game and the Shapley value. \citeauthor{b29} [\citeyear{b29}] derived an optimal contract and pricing mechanism to address the dynamic asynchronous issue. The above works focus on optimizing FL performance through optimal scheme design or clients selection algorithm, but cannot be directly adapted to address the dynamic and incomplete information context in FL. Unlike the above studies, we discuss and derive an approximate optimal mechanism that not only improves the performance but also operates under the context of TII, making it applicable to real-world FL.

\textbf{Incomplete Information in FL.}  In a realistic FL platform, the server and clients have limited information about each other in terms of available resources \cite{b30}. Some works had proposed practical scenarios for recovering this incomplete information. For instance, \citeauthor{b13} [\citeyear{b13}] and \citeauthor{b28} [\citeyear{b28}] designed the optimal contract in the presence of incomplete information about clients’ type for the server. \citeauthor{b9} [\citeyear{b9}] and \citeauthor{b32} [\citeyear{b32}] modeled a Bayesian game to make resource or reward allocation decisions with incomplete information, where a client lacks personal information about others. Existing works are mainly concerned about one-sided incomplete information scenario, where the server is unaware of clients’ inherent information, or clients’ lack information about other clients or the server. However, a practical FL platform is a dynamic changing TII context, where not only the server unknown the clients’ inherent dynamic computation resources, but also the clients are unaware of the server’s real-time communication resources, which need to be allocated to those of the clients. To address this challenge, we model the Bayesian persuasion game and further analyze the optimal signal and payment rules in the proposed mechanism under context of TII in FL.

\textbf{OFL.} OFL performs model updates immediately upon a random client arrives \cite{b6}. Most existing works in OFL focus on resource allocation and clients selection research. For instance, \citeauthor{b6}[\citeyear{b6}] and \citeauthor{b36}[\citeyear{b36}] designed update strategies and adaptive gradient sparsification techniques to reduce resource consumption. \citeauthor{b37}[\citeyear{b37}] and \citeauthor{b2}[\citeyear{b2}] addressed the client selection through staleness awareness and performance prediction. Existing works overlook the resources dynamic and incomplete information in OFL, which may not applicable in practical scenarios and prevent its successful sustainable development. In contrast, we consider these context and design an incentive mechanism for a more general and practical scenario in OFL.

\section{Problem Formulation}

In this section, we introduce the system overview, pricing model, and formulate the server’s cost minimization problem in OFL.

\subsection{System Overview of OFL}

In OFL, the global model is dynamically and sequentially updated in real-time based on clients' local training results. The OFL framework consists of a server and $N$ clients, denoted by $\mathcal{N} = \{1,\cdots,N\}$, and operates over a series of time slots $\mathcal{T} = \{ 1,\cdots,T \}$. At each time slot $t\in \mathcal{T}$, a client $n_t\in\mathcal{N}$ arrives at the OFL platform and determines whether to join the training. We define $\omega_t$ and $\omega_t^n$ as the global model parameters and the local model parameters of client $n_t$ respectively, and let $x_n^t$ denotes the local training data of client $n_t$. If the arriving client $n_t$ decides to participate in the FL training, she will perform her local update process as follows:
\begin{equation}
	\omega_{t+1}^n = \omega_t^n - \eta\triangledown l(\omega_t, x_t^n),
\end{equation}
where $\eta$ is the learning rate, and $\triangledown l(\omega_t, x_t^n)$ is the gradient of local average loss $l(\omega_t, x_t^n)$.

After the client $n_t$ finishes the local update, the server integrates the received update into the global model immediately, without waiting for all clients to complete their local computations and transmissions, i.e., 
\begin{equation}
	\label{eq2}
	\omega_{t+1} = (1-\alpha)\omega_t + \alpha\omega_{t+1}^n,
\end{equation}
where $\alpha$ is the weighting factor that controls the server’s incorporation of the client’s update into the global model.

\subsection{Pricing Model}

If the client $n_t$ decides to join the FL training, her utility is the difference between the reward $\gamma_t\in[\underline{\gamma},\overline{\gamma}]$ offered by the server and the cost $c_t$ incurred during model training. The cost $c_t$ consists of computation and communication consumption \cite{b18}. Computation cost (measured by client CPU cycles and CPU frequency) for computing local gradients \cite{b17}. Communication cost (measured by allocated bandwidth) for transmitting local gradients to the server \cite{b10}. The client $n_t$ with adequate computing resources can complete local training more efficiently by reducing computational delays and computation energy consumption. Then, the higher the computation resources $\theta_t\in[\underline{\theta},\overline{\theta}]$ that the client $n_t$ has, the lower the computation cost. The client $n_t$ with adequate communication resources can accelerate the upload of local gradients, reducing communication time and communication energy consumption \cite{b9}. Then, the higher the communication resources $\tau_t\in[\underline{\tau},\overline{\tau}]$ assigned by the server to the client $n_t$, the lower the communication cost. Therefore, the cost of client $n_t$ is determined by the computation resources $\theta_t$ inherent to the client and  the communication resources $\tau_t$ assigned by the server, i.e. $c_t(\theta_t,\tau_t)$. Note that the cost function $c_t(\theta_t,\tau_t)$ related to $\theta_t$ and $\tau_t$ is public knowledge for both the server and the clients. All clients have the same form of cost function. Therefore, we omit the subscript $t$ for $c_t$, i.e., $c_t(\theta_t,\tau_t)$. Rational clients are willing to join FL if and only if the received reward exceeds the suffered cost $c(\theta_t,\tau_t)$, i.e. $\gamma_t \ge c(\theta_t,\tau_t)$.

One round of the OFL process can be detailed as follows: First, the server sends the latest reward $\gamma_t$ and allocable available communication resources $\tau_t$ to the arriving client $n_t$ (step $\textcircled{\small 1}$). Second, the client estimates the cost $c(\theta_t,\tau_t)$ and determines whether to join the FL training process (step $\textcircled{\small 2}$). If the client decides to engage in FL training, the server sends the latest global model parameters $\omega_t$ to the client, and the client sends back the updated local model parameters $\omega_{t+1}^n$ to the server after finishing local training. Otherwise, the server needs to wait for the next client arrive and participate in training (step $\textcircled{\small 3}$). Finally, the server updates the global model $\omega_{t+1}$ according to Eq. (\ref{eq2}) immediately.

\begin{figure}
	\centering
	\includegraphics[height=3.1cm,width=8.3cm,angle=0,scale=1]{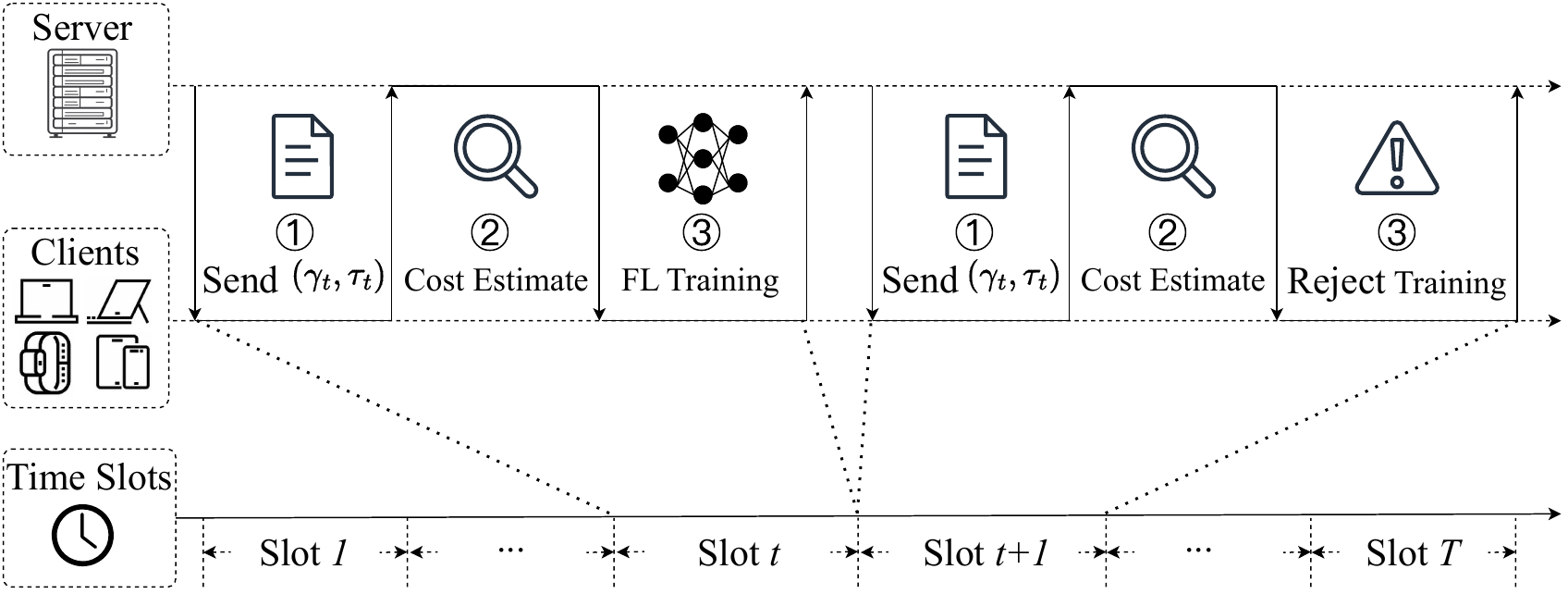}
	\caption{An overview of OFL.}
\end{figure}

\subsection{Cost Minimization Problem}

The OFL process terminates when the global model accuracy reaches a predefined threshold $\varepsilon$, supposed at the final time slot $T$ \cite{b34}. Let $v_s(\varepsilon)$ represent the actual valuation the server receives with global accuracy $\varepsilon$. Then, the server’s objective of maximizing its utility $u_s$ can be formulated as follows:
\begin{equation}
	\begin{aligned}
		&\max_{(\gamma_t,\tau_t)} u_s = \max [v_s(\varepsilon) - c_s(\gamma_t,\tau_t)],\\
		&\text{ s.t. } c_s(\gamma_t,\tau_t) = \sum_{t=1}^T \mathbf{1}[\gamma_t\ge c(\theta_t,\tau_t)]\gamma_t,
	\end{aligned}
\end{equation}
where $\mathbf{1}(\gamma_t\ge c(\theta_t,\tau_t))$ is an indicator function indicating whether client $t$ joins the FL training. We declare that the reward given to clients here as the server’s cost $c_s$, which the server needs to compensate for the clients’ service.

Considering that the global accuracy $\varepsilon$ is fixed, $v_s(\varepsilon)$ is constant and can therefore be ignored. Clients with lower computation resources will lead to training bias, effecting the model’s generalization capability and causing overfitting issues. To prevent this issue, the server needs to filter clients whose computation resources $\theta_t$ are below the threshold $\beta$ by adjusting the reward $\gamma_t$. Consequently, maximizing the server’s utility can be reformulated as follows:
\begin{equation} 
	\label{eq5}
	\begin{aligned}
		&\min_{(\gamma_t,\tau_t)} c_s(\gamma_t,\tau_t),\\
		&\text{ s.t.}
		\begin{cases}
			c_s(\gamma_t,\tau_t) = \sum_{t=1}^T \mathbf{1}[\gamma_t\ge c(\theta_t,\tau_t)]\gamma_t,\\
			\min \{ \theta_t|\gamma_t\ge c(\theta_t,\tau_t) \}\ge\beta.
		\end{cases}
	\end{aligned}
\end{equation}

In the context of TII, it is difficult for the server to minimize the cost without having complete information about each client’s decision to join the FL training. On the one hand, the communication resource $\tau_t$ assigned by the server to client $t$ is confidential to the client. On the other hand, the computation resource $\theta_t$ inherent to client $t$ is private information for the server. Therefore, it is challenging for a client to decide whether to participate in OFL training, and for the server to provide a suitable reward that ensures adequate clients join while minimizing the cost.

\section{DaringFed Mechanism}

In this section, we model the server’s cost minimization problem in OFL under TII using  the Bayesian persuasion game, and design a novel DaringFed mechanism to address it.

\subsection{Bayesian Persuasion Game}

The Bayesian persuasion (BP) game is a game-theoretic framework that investigates how the sender can strategically disclose private information through sending signals to persuade the receiver to make decisions favorable to the sender \cite{b19}. Specifically, in this model, the server acts as a sender, sending signals, while the client acts as a receiver, receiving signals and making decision. Signals $\sigma\in\Sigma$ are sent to update the posterior distribution of communication resource. We assume that the computation resource $\theta_t$ and communication resource $\tau_t$ are i.i.d. across all clients, thus, we can omit subscript $t$ for Eq. (\ref{eq5}). Therefore, the Bayesian persuasion problem can be formulated as the  expected server’s cost for any client $t$, i.e.,

\begin{equation}
	\label{eq6}
	\begin{aligned}
		&\arg \min_{(\gamma^*,\sigma^*)} c_s(\gamma,\sigma),\\
		&\text{ s.t.}
		\begin{cases}
			c_s(\gamma,\sigma) = \sum_{t=1}^T \mathbf{1}[\gamma\ge c(\theta,\sigma)]\gamma,\\
			\min \{ \theta|\gamma\ge c(\theta,\tau) \}\ge\beta,
		\end{cases}
	\end{aligned}
\end{equation}
where $\gamma^*$ and $\sigma^*$ represent the optimal reward and signal that minimize the server’s cost, respectively.

\subsection{Formalization of DaringFed Mechanism}

DaringFed integrates a Bayesian persuasion signal rule to estimate the posterior expectation of the communication resources $\tau$, and a dynamic pricing rule to determine the amount of reward $\gamma$ in a TII environment.

\begin{myDef}
	(DaringFed) DaringFed mechanism is represented as a $2$-tuple $(\mathbb{S}, \mathbb{P})$, i.e., a Bayesian persuasion signal rule $\mathbb{S}$, and a dynamic pricing rule $\mathbb{P}$.
	\begin{itemize}
		\item $\mathbb{S}$: $[\underline{\tau},\overline{\tau}]\times R^+\rightarrow R^+$ determines how the distribution of the signal $\sigma$ align with communication resources $\tau$, i.e., 
		\begin{equation}
			\rho(\sigma|\tau)\leftarrow\frac{\phi(\tau|\sigma)}{\lambda(\tau)},
		\end{equation}
		where $\phi(\cdot)$ is the Bayesian posterior distribution of $\tau$ after receiving signal $\sigma$, and $\lambda(\cdot)$ is the public prior distribution of $\tau$.
		\item $\mathbb{P}$: $R^+\times [\underline{\theta},\overline{\theta}]\rightarrow R^+$ denotes the reward that the server can provide to a client, which is determined by the signal rule and the client’s computation resources $\theta$, i.e., 
		\begin{equation}
			\gamma\leftarrow\rho(\sigma|\tau)\theta.
		\end{equation}
	\end{itemize}
\end{myDef}

We define $\mu$ as the posterior mean of $\tau$, and transfer the Bayesian persuasion signal rule in DaringFed into selecting the distribution of $\mu$ over $\tau$. And hence, Eq. (\ref{eq6}) can be rewritten as the following optimization problem:
\begin{equation}
	\begin{aligned}
		&\arg\min_{(\gamma^*,\rho^*)} c_s(\gamma,\rho),\\
		&\text{ s.t. }
		\begin{cases}
			c_s =  \gamma \int_{\underline{\tau}}^{\overline{\tau}} \lambda(\tau) \int_{\underline{\tau}}^{\overline{\tau}} \rho(\mu|\tau) \psi(\gamma,\mu) d\mu d\tau, \\
			\min \{ \theta| \gamma\ge \int_{\underline{\tau}}^{\overline{\tau}}\lambda(\tau)\int_{\underline{\tau}}^{\overline{\tau}}\rho(\mu|\tau)c(\theta,\mu)d\mu d\tau \}\ge\beta,
		\end{cases}
	\end{aligned}
\end{equation}
where $\psi(\gamma,\mu)$ defines the probability that a client is willing to join FL training when the reward is $\gamma$ and her posterior estimate of the communication resources is $\mu$. 

To ensure the posterior belief remains correct after being updated with prior public information and signals, the distribution of $\rho$ is feasible if and only if it satisfies the Bayesian Consistency (BayesCon) \cite{b8}. Additional, to guarantee the signals encourage clients to adjust their strategies in ways that benefit the server, while maintaining alignment between the clients’ expected posterior belief and the prior, the distribution of $\rho$ should also satisfy Bayesian Plausible (BayesPla) and Bayesian Benefit (BayesBen) \cite{b19}. We provide formal definitions of BayesCon, BayesPla, and BayesBen as follows.

\begin{myDef}
	(BayesCon) $\rho$ satisfies BayesCon if
	\begin{equation}
		\label{eq7}
		\frac{\int_{\underline{\tau}}^{\overline{\tau}} \lambda(\tau)\rho(\mu|\tau)\tau d\tau}{\int_{\underline{\tau}}^{\overline{\tau}} \lambda(\tau)\rho(\mu|\tau) d\tau} = \mu.
	\end{equation}
\end{myDef}

\begin{myDef}
	(BayesPla) $\rho$ satisfies BayesPla if 
	\begin{equation}
		\label{eq8}
		\int_{\underline{\tau}}^{\overline{\tau}} \lambda(\tau) \int_{\underline{\tau}}^{\overline{\tau}} \rho(\mu|\tau) \mu d\mu d\tau = \int_{\underline{\tau}}^{\overline{\tau}} \lambda(\tau) \tau d\tau.
	\end{equation}
\end{myDef}

\begin{myDef}
	(BayesBen) $\rho$ satisfies BayesBen if
	\begin{equation}
		\label{eq9}
		\int_{\underline{\tau}}^{\overline{\tau}} \lambda(\tau) \int_{\underline{\tau}}^{\overline{\tau}} \rho(\mu|\tau) \psi(\gamma,\mu) d\mu d\tau \ge  \int_{\underline{\tau}}^{\overline{\tau}} \lambda(\tau) \psi(\gamma,\tau)d\tau.
	\end{equation}
\end{myDef}

Therefore, the DaringFed mechanism design problem can be formulated as:
\begin{equation} 
	\label{eq10}
	\begin{aligned}
		&\arg \min_{(\gamma^*,\rho^*)} c_s(\gamma,\rho),\\
		&\text{ s.t.}
		\begin{cases}
			c_s(\gamma,\rho) =  \gamma \int_{\underline{\tau}}^{\overline{\tau}} \lambda(\tau) \int_{\underline{\tau}}^{\overline{\tau}} \rho(\mu|\tau) \psi(\gamma,\mu) d\mu d\tau,\\
			\min \{ \theta| \gamma\ge \int_{\underline{\tau}}^{\overline{\tau}}\lambda(\tau)\int_{\underline{\tau}}^{\overline{\tau}}\rho(\mu|\tau)c(\theta,\mu)d\mu d\tau \}\ge\beta,\\
			\frac{\int_{\underline{\tau}}^{\overline{\tau}} \lambda(\tau)\rho(\mu|\tau)\tau d\tau}{\int_{\underline{\tau}}^{\overline{\tau}} \lambda(\tau)\rho(\mu|\tau) d\tau} = \mu,\\
			\int_{\underline{\tau}}^{\overline{\tau}} \lambda(\tau) \int_{\underline{\tau}}^{\overline{\tau}} \rho(\mu|\tau) \mu d\mu d\tau = \int_{\underline{\tau}}^{\overline{\tau}} \lambda(\tau) \tau d\tau,\\
			\int_{\underline{\tau}}^{\overline{\tau}} \lambda(\tau) \int_{\underline{\tau}}^{\overline{\tau}} \rho(\mu|\tau) \psi(\gamma,\mu) d\mu d\tau \ge  \int_{\underline{\tau}}^{\overline{\tau}} \lambda(\tau) \psi(\gamma,\tau)d\tau.
		\end{cases}
	\end{aligned}
\end{equation}
DaringFed aims to minimize the server’s cost, while ensuring clients‘ computation resources exceed a threshold, and satisfies BayesCon, BayesPla and BayesBen constraint under the optimal reward $\gamma^*$ and signal $\rho^*$. Analyzing the existence of $\gamma^*$ and $\rho^*$, and further identifying the optimal $\gamma^*$ and $\rho^*$, constitutes the objective of DaringFed.

\section{Design of DaringFed Mechanism}

Due to the use of upper confidence bound for estimating the distribution of computation resources among clients, which involves a discrete decision space and limited exploration time, we can only derive an approximate optimal DaringFed. Consequently, in this section, we first analyze the existence of optimal DaringFed mechanism under one-sided incomplete information, where the server knows the distribution of computation resources among clients, and then design an approximately optimal DaringFed mechanism under a TII scenario.

\subsection{Existence of Optimal DaringFed Mechanism}

We define that $\theta$ follows a distribution with survival function $s(\theta)$, describing the probability that the computation resources exceed a threshold $\theta$. Recall the indicator function $\mathbf{1}(\gamma\ge c(\theta,\mu))$, which indicates whether a client decides to join the FL. The values of $\gamma$ and $\mu$ determine the minimum threshold for a client’s computation resources $\hat{\theta}$, i.e. $\hat{\theta} = \min \{\theta | \mathbf{1}(\gamma\ge c(\theta,\mu))\}$. If the client’s computation resources $\theta$ exceeds this threshold $\hat{\theta}$, the client is willing to join FL. Following the survival distribution $s(\theta)$, the probability that a client’s computation resources exceeds the threshold $\hat{\theta}$ is $s(\hat{\theta}) = s( \min \{\theta | \mathbf{1}(\gamma\ge c(\theta,\mu))\} )$, which can be simplified as $s(\gamma,\mu)$. Therefore, we can reformulate Eq. (\ref{eq10}) as:
\begin{equation} 
	\label{eq15}
	\begin{aligned}
		&\arg \min_{(\gamma^*,\rho^*)} c_s(\gamma,\rho),\\
		&\text{ s.t.}
		\begin{cases}
			c_s(\gamma,\rho) =  \gamma \int_{\underline{\tau}}^{\overline{\tau}} \lambda(\tau) \int_{\underline{\tau}}^{\overline{\tau}} \rho(\mu|\tau) s(\gamma,\mu) d\mu d\tau,\\
			\min \{ \theta| \gamma\ge \int_{\underline{\tau}}^{\overline{\tau}}\lambda(\tau)\int_{\underline{\tau}}^{\overline{\tau}}\rho(\mu|\tau)c(\theta,\mu)d\mu d\tau \}\ge\beta,\\
			\frac{\int_{\underline{\tau}}^{\overline{\tau}} \lambda(\tau)\rho(\mu|\tau)\tau d\tau}{\int_{\underline{\tau}}^{\overline{\tau}} \lambda(\tau)\rho(\mu|\tau) d\tau} = \mu,\\
			\int_{\underline{\tau}}^{\overline{\tau}} \lambda(\tau) \int_{\underline{\tau}}^{\overline{\tau}} \rho(\mu|\tau) \mu d\mu d\tau = \int_{\underline{\tau}}^{\overline{\tau}} \lambda(\tau) \tau d\tau,\\
			\int_{\underline{\tau}}^{\overline{\tau}} \lambda(\tau) \int_{\underline{\tau}}^{\overline{\tau}} \rho(\mu|\tau) s(\gamma,\mu) d\mu d\tau \ge  \int_{\underline{\tau}}^{\overline{\tau}} \lambda(\tau) s(\gamma,\tau)d\tau.
		\end{cases}
	\end{aligned}
\end{equation}

Our goal in designing DaringFed is to establish an ideal Bayesian persuasion Nash equilibrium (BPNE) when $s(\gamma,\mu)$ is known to the server, with the objective of maximizing the server’s utility. To analyze the equilibrium of the BP game, we make the following common assumptions regarding the cost function $c(\theta,\tau)$ and survival function $s(\theta)$.

\begin{myAss}
	The client’s cost function $c(\theta,\tau)$ satisfies:
	\begin{itemize}
		\item Given any $\tau$, the function $c(\theta,\tau)$ is a non-increasing, convex function w.r.t. $\theta$.
		\item Given any $\theta$, the function $c(\theta,\tau)$ is a non-increasing, convex function w.r.t. $\tau$.
	\end{itemize}
\end{myAss}

The assumption on the client’s cost function are standard in many economics scenarios \cite{b20,b21}. It is natural to assume that a client’s cost decrease with its ability and the communication resource. The convexity of cost functions is commonly used to reflect increasing marginal costs.

\begin{myAss}
	The survival function $s(\theta)$ is a non-increasing, concave function w.r.t. $\theta$.
\end{myAss}

The assumption regarding the survival function has ample justifications for it being a non-increasing, concave function \cite{b9,b22}. It is reasonable to assume that as the threshold value of $\theta$ increases, the number of clients who meet this criterion decreases, leading to a lower probability. The concave of the survival function is generally used to capture decreasing marginal probability.

In the BPNE, the server selects a distribution $\rho$ and a reward $\gamma$ that maximize its utility, as defined below.

\begin{myDef}
	(BPNE) Let $(\gamma^*,\rho^*)$ denotes the BPNE, i.e.,
	\begin{equation}
		\begin{cases}
			c_s(\gamma^*,\rho^*) \ge c_s(\gamma^*,\rho), \\
			c_s(\gamma^*,\rho^*) \ge c_s(\gamma,\rho^*). \\
		\end{cases}
	\end{equation}
\end{myDef}

At the BPNE, the server cannot benefit from changing its signals or rewards. Then, we investigate the existence and uniqueness of a BPNE.

\begin{myLem}
	There exists a unique BPNE in the BP game.
\end{myLem}

Then, we derive the unique BPNE satisfying BayesCon, BayesPla, and BayesBen.

\begin{myThe}
	The optimal signals $\rho^*$ over posterior means $\mu$ for the given prior value $\tau\in[\underline{\tau},\bar{\tau}]$ are
	\begin{equation}
		\begin{cases}
			\rho(\mu|\underline{\tau})=\frac{\rho(\mu)(\overline{\tau}-\mu)}{\lambda(\underline{\tau})(\overline{\tau} - \underline{\tau})},\\
			\rho(\mu|\overline{\tau})=\frac{\rho(\mu)(\mu-\underline{\tau})}{\lambda(\overline{\tau})(\overline{\tau} - \underline{\tau})},
		\end{cases}
	\end{equation}
	where $\rho(\mu)= \frac{ \lambda(\underline{\tau})\underline{\tau} + \lambda(\bar{\tau})\bar{\tau} }{\mu}$, and the optimal reward $\gamma^*$ is $\arg\min\gamma\int_{\underline{\tau}}^{\overline{\tau}}\rho(\mu)s(\gamma,\mu)d\mu$.
\end{myThe}

\subsection{Approximate Optimal DaringFed Mechanism under TII}

In this section, we analyze the approximate optimal DaringFed mechanism in a TII scenario. Due to the uncertainty of $s(\gamma,\mu)$, the server needs to estimate it with respect to $\gamma$ and $\mu$ first, and then design the optimal signal $\rho$ and $\gamma$ to maximize its expected utility.

To estimate $s(\cdot)$, we considering the reward space and the computation resources space are discrete spaces, i.e., $\gamma\in R$ and $\theta\in\Theta$, where $R = \{\underline{\gamma}, \underline{\gamma}+\xi,\cdots,\overline{\gamma}-\xi,\overline{\gamma}\}$ and $\Theta=\{\underline{\theta},\underline{\theta}+\xi,\cdots,\overline{\theta}-\xi,\overline{\theta}\}$. Let $\mathcal{N}_t(\hat{\theta})$ represents the set of time rounds prior to time slot $t$ where the threshold equals $\hat{\theta}$, i.e., $\mathcal{N}_t(\hat{\theta}) = \{ n| (n\le t) \cap (\hat{\theta} = \min \{\theta | \mathbf{1}(\gamma\ge c(\theta,\mu))\}) \}$. $N_t(\hat{\theta})$ denote the total number of such time slots, i.e., $N_t(\hat{\theta}) = |\mathcal{N}_t(\hat{\theta})|$. Then, we can derive $\overline{s}(\hat{\theta})$ as follows.

\begin{myLem}
	The upper confidence bound for the probability that the computation resources threshold is exactly $\hat{\theta}$ is:
	\begin{equation}
		\label{eq16}
		\overline{s}(\hat{\theta}) = \frac{\sum_{n\in\mathcal{N}_t(\hat{\theta})}\mathbf{1}(\gamma_n \ge c(\theta_n,\mu_n))}{{N_t(\hat{\theta})}} + \sqrt{\frac{\ln N}{2N_t(\hat{\theta})}}.
	\end{equation}
\end{myLem}

We define the approximate optimal reward as $\gamma^+\in R$, and the corresponding threshold for clients’ computation resources as $\hat{\theta}^+ = \min\{ \theta|\mathbf{1}(\gamma^+\ge c(\theta,\mu)) \}$. There exists $z\in\mathbb{Z}$ such that $(z-1)\xi\le\hat{\theta}^+\le z\xi$ for $\theta\in\Theta$. Then, we can find two posterior expectation of communication resources, $\mu_r$ and $\mu_l$, such that $(z-1)\xi = \min \{\theta | \mathbf{1}(\gamma^+ \ge c(\theta,\mu_r))\})$ and $z\xi = \min \{\theta | \mathbf{1}(\gamma^+\ge c(\theta,\mu_l))\})$. In the following, we derive the approximate optimal signals $\rho^+$ from a given approximate optimal reward $\gamma^+\in R$ using Theorem 2, and then design Algorithm 1 to obtain both $\gamma^+$ and $\rho^+$ globally.

\begin{myThe}
	The approximate optimal signals $\rho^+$ that ensures the clients’ computation resources threshold $\hat{\theta}$ exactly at the discrete space $\Theta$ are
	\begin{equation}
		\label{eq17}
		\begin{cases}
			\rho^+(\mu_l|\underline{\tau})=\frac{\rho(\mu_l)(\overline{\tau}-\mu_l)}{\lambda(\underline{\tau})(\overline{\tau} - \underline{\tau})},\\
			\rho^+(\mu_l|\overline{\tau})=\frac{\rho(\mu_l)(\mu_l-\underline{\tau})}{\lambda(\overline{\tau})(\overline{\tau} - \underline{\tau})},\\
			\rho^+(\mu_r|\underline{\tau})=\frac{\rho(\mu_r)(\overline{\tau}-\mu_r)}{\lambda(\underline{\tau})(\overline{\tau} - \underline{\tau})},\\
			\rho^+(\mu_r|\overline{\tau})=\frac{\rho(\mu_r)(\mu_r-\underline{\tau})}{\lambda(\overline{\tau})(\overline{\tau} - \underline{\tau})},\\
		\end{cases}
	\end{equation}
	where $\rho(\mu_l) = \frac{\rho(\mu)(\mu_r-\mu)}{\mu_r-\mu_l}$, and $\rho(\mu_r) = \frac{\rho(\mu)(\mu-\mu_l)}{\mu_r-\mu_l}$.
\end{myThe}

\begin{algorithm}[bt]
	\caption{Approximate Optimal Design of DaringFed}
	\KwIn{$\xi$, $\Theta = \{\underline{\theta}, \underline{\theta}+\xi,\cdots,\overline{\theta}-\xi,\overline{\theta}\}$, $R = \{ \underline{\gamma},\underline{\gamma}+\xi,\cdots,\overline{\gamma}-\xi,\overline{\gamma} \}$, $\underline{\tau}$, $\overline{\tau}$, $\tau$;}
	\KwOut{$\rho$, $\gamma$;}
	
	\For{$t\in \{1,2,\cdots,|\Theta|\}$}
	{
		$\theta \leftarrow \underline{\theta}+t(\xi-1)$\;
		$\gamma \leftarrow \{\theta = \min\{ \theta|\mathbf{1}(\gamma\ge c(\theta,\tau)) \}\}$\;
	}
	
	\For{$t = |\Theta| + 1; t< T; t++$}
	{
		Traverse $\theta\in\Theta$ to update $s(\theta)$ by Eq. (\ref{eq16})\;
		\For{$\gamma\in R$}
		{
			$z \leftarrow \{z | z\in\mathbb{Z}^+ \}\cap \{ (z-1)\xi\le\theta\le z\xi \}$\;
			$\mu_r \leftarrow \{ (z-1)\xi = \min\{ \theta|\mathbf{1}(\gamma\ge c(\theta,\mu_r)) \} \}$\;
			$\mu_l \leftarrow \{ z\xi = \min\{ \theta|\mathbf{1}(\gamma\ge c(\theta,\mu_l)) \} \}$\;
			Obtain $\rho(\mu_l|\underline{\tau})$, $\rho(\mu_l|\overline{\tau})$, $\rho(\mu_r|\underline{\tau})$, and $\rho(\mu_r|\overline{\tau})$ by Eq. (\ref{eq17})\;
			Obtain $c_s$ by Eq. (\ref{eq15})\;
			Retain $\gamma$ and $\rho$ that minimizes $c_s$.
		}
	}
\end{algorithm}

In Algorithm 1, to initialize $s(\theta)$, we traverse each element $\theta\in \Theta$ where $\theta=\hat{\theta}$, without any signaling mechanism, i.e., the posterior expectation of communication resources $\mu$ is equal to real value $\tau$ directly (lines 1-3). Then, for each arriving client, we update $s(\theta)$ based on the historically information, where the provided reward $\gamma$ and signals $\rho$ can persuasion clients to participate in FL training (line 5). Finally, we traverse $\gamma\in R$ to find the suitable $\gamma$ and corresponding $\rho$ according to Theorem 2, in order to minimize the server’s cost $c_s$ (lines 6-12). The retained $\gamma$ and $\rho$ represent the selected reward and signals for the current time slot, and this process can iterate continuously until $\gamma$ convergence to $\gamma^+$, or the distribution of clients’ local computation resources changes. 

In the TII scenario, the approximate optimal signals $\rho$ and reward $\gamma$ can only be obtained from the discrete space $\Theta$ and $R$, respectively. However, the optimal value may not lie within these discrete spaces. Here, we analyze the bounds on the approximate optimal value as follows.

\begin{myThe}
	\label{the3}
	The bound on the approximate optimal value is 
	\begin{equation}
		c_s(\gamma^+,\rho^+) - c_s(\gamma^*,\rho^*) \le 2\xi.
	\end{equation}
\end{myThe}

According to Theorem \ref{the3}, the approximate optimal solution obtained by Algorithm 1 in the TII almost converges to the optimal solution, with the gap bounded by $2\xi$.

\begin{figure*}
	\begin{minipage}{0.24\textwidth}
		\centering
		\includegraphics[height=3.5cm,width=4.5cm,angle=0,scale=1]{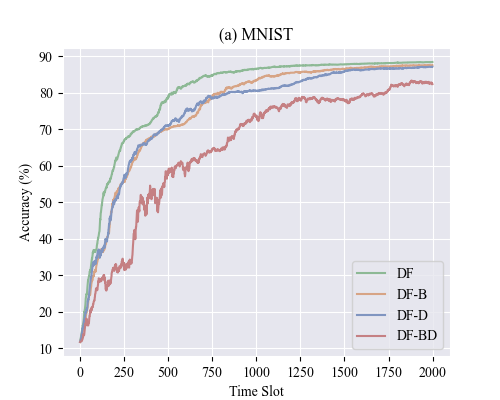}
	\end{minipage}
	\begin{minipage}{0.24\textwidth}
		\centering
		\includegraphics[height=3.5cm,width=4.5cm,angle=0,scale=1]{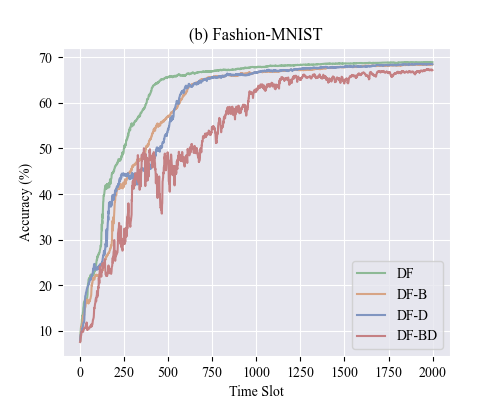}
	\end{minipage}
	\begin{minipage}{0.24\textwidth}
		\centering
		\includegraphics[height=3.5cm,width=4.5cm,angle=0,scale=1]{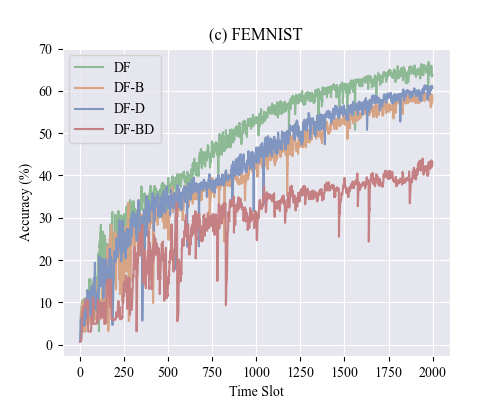}
	\end{minipage}
	\begin{minipage}{0.24\textwidth}
		\centering
		\includegraphics[height=3.5cm,width=4.5cm,angle=0,scale=1]{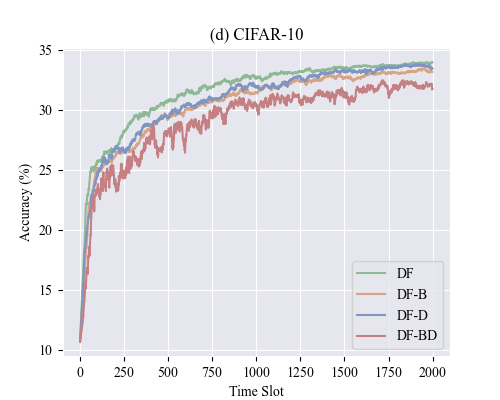}
	\end{minipage}
	\caption{ Testing the accuracy of OFL with the proposed DaringFed on (a) MNIST, (b) Fashion-MNIST, (c) FEMNIST, and (d) CIFAR-10. }
\end{figure*}

\begin{figure}[h]
	\centering
	\begin{minipage}{0.23\textwidth}
		\centering
		\includegraphics[height=3.5cm,width=4.5cm,angle=0,scale=1]{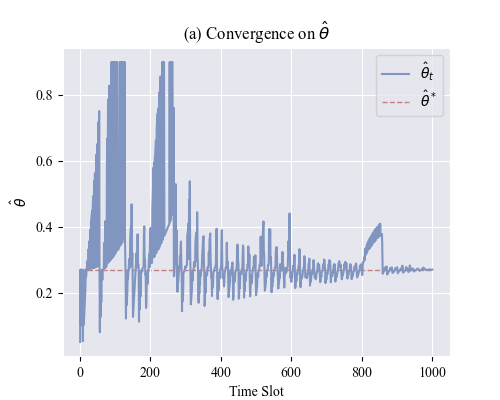}
	\end{minipage}
	\hfill
	\begin{minipage}{0.23\textwidth}
		\centering
		\includegraphics[height=3.5cm,width=4.5cm,angle=0,scale=1]{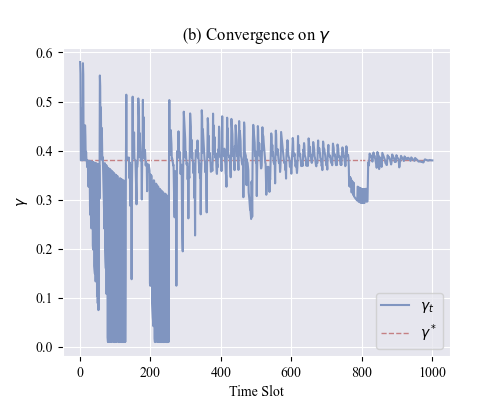}
	\end{minipage}
	\caption{ Convergence on (a) $\hat{\theta}$ and (b) $\gamma$. }
\end{figure}

\section{Experiments}

In this section, we utilize four real datasets to evaluate the performance of the DaringFed under varying level of computation resources $\theta$ and communication resources $\tau$. Subsequently, we employ a synthetic dataset to demonstrate the effectiveness of the DaringFed in terms of the convergence of the computation resources threshold $\hat{\theta}$ and the reward $\gamma$. Furthermore, we compare the improvements achieved by the DaringFed with non-DaringFed in relation to the computation resources threshold $\hat{\theta}$ and server’s cost $c_s$.

\textbf{Setup on real datasets.} We estimate the performance of the DaringFed on four real-world datasets: MNIST \cite{b24}, Fashion-MNIST \cite{b25}, FEMNIST \cite{b26}, and CIFAR-10 \cite{b27}. For MNIST and Fashion-MNIST, we employ a 4-layer convolutional neural network (CNN) model consisting of two $5\times5$ convolutional layers with 10 and 20 channels, respectively. For FEMNIST, we employ a 3-layer CNN model, consisting of a $7\times7$ convolutional layer with 32 channels, a $3\times3$ convolution layer with 64 channels, and a fully connected layer. For CIFAR-10, we employ a 5-layer CNN model consisting of two $5\times5$ convolution layers , each with 64 channels, followed by ReLU activation and $3\times3$ max pooling, and three fully connected layers. All of the above models are trained for 2000 time slots. Note that in the OFL platform, only one client arrives per time slot.

\textbf{Setup on synthetic datasets.} We analyze the impact of convergence on $\hat{\theta}$ and $\gamma$, as well as the improvements in $\hat{\theta}$ and $c_s$ using a synthetic dataset. We generate $\tau$ from a uniform distribution over the interval $[0.1, 0.9]$. We define $c=(1-\theta)^2(1.2-\mu)^2$, which satisfies Assumption 1, where $\mu,\theta\in[0.1,0.9]$. Additionally, we define $s(\theta)=1-(\frac{\theta-0.1}{0.8})^8$, where $\theta\in[0.1,0.9]$, to satisfy Assumption 2. For the discrete space in a TII scenario, we specific $\xi$ as 0.01.

\textbf{Performance of DaringFed.} To the best of our knowledge, DaringFed represents the initial effort in designing an incentive mechanism for OFL under TII. To showcase the effectiveness of DaringFed (DF), we conduct a comparative analysis with its variants against a range of  baselines to access the influence of different components within our proposed mechanism: (1) DaringFed (DF-B): without the Bayesian persuasion signal rule, where clients can only estimate communication resources based on the prior distribution. (2) DaringFed (DF-D): without the dynamic pricing rule, where the server provides clients with a fixed reward. (3) DaringFed (DF-BD): without both above rules. 

Figure 2 illustrates the model accuracy versus time slot for different baselines. DaringFed achieves the highest accuracy and converges the fastest as shown in Table 1. DF-B and DF-D exhibit lower accuracy and slower convergence, while DF-BD has the worst accuracy and the slowest convergence. This demonstrates that DaringFed improves model performance by filtering out clients with low computation or communication resources in OFL. This is because only clients with sufficient computation or communication resources can meet the constraint that the cost is less than reward under a specific reward threshold, enabling their participation in OFL training. Allocating an appropriate reward through DaringFed, based on the distribution of clients’ resources, can improve system performance. Even though clients resources change dynamically, DaringFed can adjust the reward based on feedback regarding whether each client join FL during per time slot.

\begin{table}[t!]
	\centering
	\begin{tabular}{lcccc}
		\toprule
		\multirow{2}{*}{Datasets} &  \multicolumn{4}{c}{Accuracy ($\%$)}  \\
		\cmidrule(r){2-5} 
		& DF & DF-B & DF-D & DF-BD  \\
		\midrule
		MNIST & \textbf{88.38} & 87.56 & 86.93 & 83.35   \\
		Fashion-MNIST & \textbf{69.11} & 68.65 & 68.14 & 67.18 \\
		FEMNIST & \textbf{64.41} & 57.41 & 60.35   & 43.05 \\
		CIFAR-10 & \textbf{34.73}  & 34.27 & 34.48 & 31.73  \\
		\bottomrule
	\end{tabular}
	\caption{Accuracy among different thresholds.}
	\label{tab:results}
\end{table}

\begin{figure}[h]
	\centering
	\begin{minipage}{0.23\textwidth}
		\centering
		\includegraphics[height=3.5cm,width=4.5cm,angle=0,scale=1]{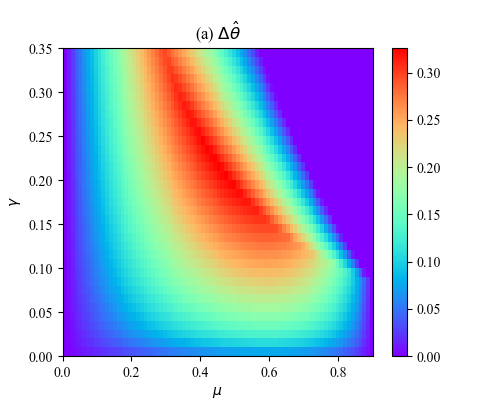}
	\end{minipage}
	\hfill
	\begin{minipage}{0.23\textwidth}
		\centering
		\includegraphics[height=3.5cm,width=4.5cm,angle=0,scale=1]{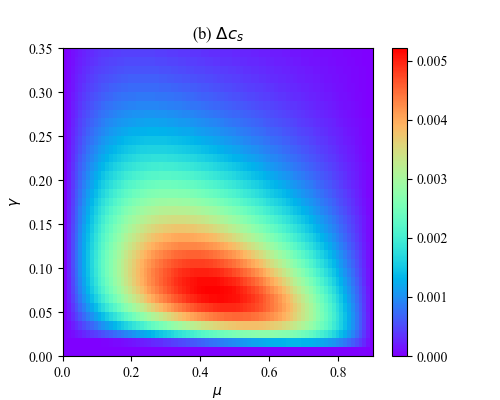}
	\end{minipage}
	\caption{ Improvements on (a) $\hat{\theta}$ and (b) $c_s$. }
\end{figure}
  
\textbf{Impact of Convergence on $\hat{\theta}$ and $\gamma$.} Figure 3 shows that the client’s local computation resource threshold $\hat{\theta}$ and reward $\gamma$ can converge after numerous time slot iterations. Each time slot has an optimal computation resource threshold $\hat{\theta}^*$, which is determined by the arriving client’s $\theta$. The optimal threshold $\hat{\theta}^*$ ensures reward equal to client’s cost exactly, thereby maximizing the server’s utility. However, $\hat{\theta}^*$ is unknown to the server, as the server is unaware of the distribution of the client’s computation resource. There are multiple discrete values of $\hat{\theta}$ for the server to select from. The server needs to evaluate $\hat{\theta}$ by exploiting the available known knowledge that the selected $\hat{\theta}$ can persuade clients to join FL training, and by exploring the unknown knowledge regarding the unselected $\hat{\theta}$, which has never been chosen before. The server can estimate the distribution of $\theta$ through above process, and further converge to $\hat{\theta}^*$ and $\gamma^*$. This mechanism is still applicable in scenarios where the distribution of clients’ computation resources is dynamic changing. After the distribution of clients’ computation resources changes, the convergence value may changed after training, and multiple different convergence values can exist depending on distribution.

\textbf{Improvements in $\hat{\theta}$ and $c_s$.} Figure 4 uses a heatmap to evaluate the improvement in $\hat{\theta}$ and $c_s$ from DF-BD to DaringFed across various values of $\mu$ and $\gamma$ in OFL, i.e. $\Delta\hat{\theta}=\hat{\theta}^{*}-\hat{\theta}’$ and $\Delta c_s=c_s^{*}-c_s’$, where $\cdot^{*}$ and $\cdot’$ represents the results derived from DaringFed and DF-BD, respectively. As shown in Figure 4 (a), $\Delta\hat{\theta}$ attains its maximum value for specific  values of $\gamma$ and $\mu$. On the one hand, a higher $\mu$ reduce clients’ cost, leading to lower values of $\hat{\theta}^{bp}$ and $\hat{\theta}^{non-bp}$ under a fixed $\gamma$. Since the function of $\hat{\theta}$ is a convex function according to Appendix A, when the posterior expectation of $\mu$ is fixed to satisfy Bayesian Plausible, there exists a maximum value of $\Delta\hat{\theta}$, and $\hat{\theta}^{bp}$ is greater than $\hat{\theta}^{non-bp}$, filter better clients under DaringFed. On the other hand, a higher $\gamma$ reduce clients’ cost, leading to lower values of $\hat{\theta}^{bp}$ and $\hat{\theta}^{non-bp}$ under a fixed $\mu$. $\Delta\hat{\theta}$ reaches its maximum value when $\gamma$ at the optimal value. As shown in Figure 4 (b), $\Delta c_s$ attains its maximum value for specific values of $\gamma$ and $\mu$. On the one hand, the function of $c_s$ is a concave according to Appendix A, under a fixed $\gamma$. There exists a maximum value for $\Delta c_s$ when posterior expectation of $\mu$ is fixed to satisfy Bayesian Plausible, and $c_s^{non-bp}$ is worse than $c_s^{bp}$. On the other hand, a higher $\gamma$ leads to higher values of $c_s^{bp}$ and $c_s^{non-bp}$ under a fixed $\mu$. And $\Delta c_s$ reach a maximum value when $\gamma$ is optimal.

\section{Conclusion}

In this paper, DaringFed was proposed to address the context of TII in the  OFL platform, which integrated a Bayesian persuasion signal rule and a dynamic pricing rule. To overcome the challenge that clients’ were unaware of the communication resources allocated by the server, a Bayesian persuasion signal rule was used to estimate the posterior expectation of communication resources on the clients’ side. To tackle the challenge that the server was unaware of clients’ inherent computation resources, a dynamic pricing rule was designed to incentive clients to participate in OFL under TII as much as possible to maximize the server’s utility. Finally, extensive experiments were conducted on real and synthetic dataset to validate the effectiveness of DaringFed mechanism.

\section*{Acknowledgments}

This work was supported in part by the National Natural Science Foundation of China (No. 62072411, 62372343, 62325304, 62402352, U22B2046), in part by the Key Research and Development Program of Hubei Province (No. 2023BEB024), and in part by the Open Fund of Key Laboratory of Social Computing and Cognitive Intelligence (Dalian University of Technology), Ministry of Education (No. SCCI2024TB02).

\section*{Appendix}
\section*{Proof of Lemma 1}

Let $\hat{\theta}$ be the minimum value of the local computation resources $\theta$ that satisfies the condition for a given reward reward $\gamma$ and communication resources $\mu$, i.e., $\hat{\theta} = \min\{\theta\in\Theta:c(\theta,\mu)\le\gamma\}$, and given any $\mu_1, \mu_2 \in [\underline{\tau},\overline{\tau}]$, we can drive the corresponding $\hat{\theta}_1$ and $\hat{\theta}_2$, respectively:
\begin{equation}
	\begin{cases}
		\hat{\theta}_1 = \min \{ \theta\in\Theta:c(\theta,\mu_1) \le \gamma \},\\
		\hat{\theta}_2 = \min \{ \theta\in\Theta:c(\theta,\mu_2) \le \gamma \}.\\
	\end{cases}
\end{equation}
According to the definition of $\hat{\theta}$, the following inequalities hold:
\begin{equation}
	\begin{cases}
		c(\hat{\theta}_1,\mu_1) \le \gamma, \\
		c(\hat{\theta}_2,\mu_2) \le \gamma. \\
	\end{cases}
\end{equation}

According to Assumption 1, where $c(\theta,\tau)$ is a convex function with respect to both $\theta$ and $\tau$, i.e., $c(\lambda\hat{\theta}_1 + (1-\lambda)\hat{\theta}_2,\lambda \mu_1+(1-\lambda)\mu_2)\le\lambda c(\hat{\theta}_1,\mu_1)+(1-\lambda)c(\hat{\theta}_2,\mu_2)$, we can derive the following inequality:
\begin{equation}
	c(\lambda\hat{\theta}_1 + (1-\lambda)\hat{\theta}_2,\lambda \mu_1+(1-\lambda)\mu_2)\le \gamma,
\end{equation}
where $\lambda\in[0,1]$. Therefore, $\lambda\hat{\theta}_1 + (1-\lambda)\hat{\theta}_2$ is a feasible solution to the constraint $c(\theta,\lambda \mu_1+(1-\lambda)\mu_2) \le \gamma$. Since $\hat{\theta}$ is the minimum value that satisfies constraint $\{\theta\in\Theta:c(\theta,\mu)\le\gamma\}$, we have
\begin{equation}
	\begin{split}
		&\min\{\theta\in\Theta: c(\theta,\lambda \mu_1+(1-\lambda)\mu_2)\le\gamma\} \\&
		\hspace{2em} \le \lambda\hat{\theta}_1 + (1-\lambda)\hat{\theta}_2 \\&
		\hspace{2em} \le \lambda \min \{ \theta\in\Theta:c(\theta,\mu_1) \le \gamma \} \\&
		\hspace{2em} + (1-\lambda)\min \{ \theta\in\Theta:c(\theta,\mu_2) \le \gamma \}.
	\end{split}
\end{equation}
Therefore, $\min\{\theta\in\Theta: c(\theta,\lambda \mu_1+(1-\lambda)\mu_2)\le\gamma\}$ is a convex function with respect to $\mu$ [Cui \textit{et al.}, 2023; Boyd, 2004]. Since $s(\hat{\theta})$ is a concave function with respect to $\hat{\theta}$, as defined in Assumption 2, and $\hat{\theta}$ is determined by $\gamma$ and $\mu$. We can conclude that $s(\gamma,\mu)$ is also a concave function with respect to $\mu$. 

With a fixed $\gamma$, we simplify $s(\gamma,\mu)$ as $s(\mu)$, and define $\textit{conv}(s)$ as the convex hull of the graph of $s(\mu)$, and $\underline{s}(\mu)$ is the minimum value in $\textit{conv}(s)$, i.e., $\underline{s}(\mu) = \sup \{ s(\mu) | (\mu, s) \in \textit{conv}(\mu) \}$. For each point in $\textit{conv}(s)$, i.e., $(\mu, s) \in \textit{conv}(\mu)$, there exists a posterior distribution such that $\mu$ and $s$ equal the expected values of the posterior belief, respectively. Given a prior value $\mu$, the minimum value inside the convex hull is $\underline{s}(\mu)$. Therefore, there exist optimal signals $\rho$, and the value of optimal signal lies on  $\underline{s}(\mu)$.

With a fixed $\rho$, there exists an optimal reward $\gamma$ subject to the computation resource threshold $\beta$, based on the monotonicity and continuity properties of the cost function with respect to $\gamma$. As a result, the proof is completed.

\section*{Proof of Theorem 1}

According to Lemma 1, there exists an optimal signal $\rho$ associated with specific posterior beliefs which satisfy Bayesian plausible condition. The signal minimize the server’s cost within the convex hull $\textit{conv}(\mu)$, and the optimal signals lies on $\underline{s}(\mu)$. Any point within the convex hull $\textit{conv}(\mu)$ can be represented as a linear combinations of any set of points within the convex hull. The optimal signal structure can be achieved through the two extreme point. This is because the two extreme points represent the potential boundary cases, and any point within the convex hull can be determined as a convex combination of these boundary cases. According to the second item of Assumption 2, the optimal signals lies at $\underline{\tau}$ and $\overline{\tau}$, where $\tau\in[\underline{\tau}, \overline{\tau}]$.

Based on the Bayesian Consistency and the Bayesian plausible defined in Definitions 2 and 3, respectively, we can derive the following equations from the above analysis of the optimal signals:

\begin{equation}
	\label{eq23}
	\begin{cases}
		\int_0^1\lambda(\underline{\tau}) \rho(\mu|\underline{\tau})\mu + \lambda(\bar{\tau})\rho(\mu|\bar{\tau})\mu d\mu = \lambda(\underline{\tau})\underline{\tau} + \lambda(\bar{\tau})\bar{\tau},\\
		\frac{\lambda(\underline{\tau})\rho(\mu|\underline{\tau})\underline{\tau} + \lambda(\bar{\tau})\rho(\mu|\bar{\tau})\bar{\tau}}{\lambda(\underline{\tau})\rho(\mu|\underline{\tau}) + \lambda(\bar{\tau})\rho(\mu|\bar{\tau})}=\mu.\\
	\end{cases}
\end{equation}
Based on the first sub formula in Eq. (\ref{eq23}), we can obtain  $\lambda(\underline{\tau}) \rho(\mu|\underline{\tau}) + \lambda(\bar{\tau})\rho(\mu|\bar{\tau}) = \frac{ \lambda(\underline{\tau})\underline{\tau} + \lambda(\bar{\tau})\bar{\tau} }{\mu}$. Then, we define $\rho(\mu) = \lambda(\underline{\tau}) \rho(\mu|\underline{\tau}) + \lambda(\bar{\tau})\rho(\mu|\bar{\tau})$ , and derive the optimal signals as follows:
\begin{equation}
	\begin{cases}
		\rho(\mu|\underline{\tau})=\frac{\rho(\mu)(\overline{\tau}-\mu)}{\lambda(\underline{\tau})(\overline{\tau} - \underline{\tau})},\\
		\rho(\mu|\overline{\tau})=\frac{\rho(\mu)(\mu-\underline{\tau})}{\lambda(\overline{\tau})(\overline{\tau} - \underline{\tau})}.
	\end{cases}
\end{equation}

Under the optimal signals, we can derive that the server’s cost function is 

\begin{equation}
	\begin{split}
		&c_s(\gamma, \rho^*) =\gamma [\lambda(\underline{\tau})\int_0^1\rho(\mu|\underline{\tau}) p(\gamma,\mu) d\mu \\& 
		\hspace{2em}+ \lambda(\overline{\tau})\int_0^1\rho(\mu|\overline{\tau}) p(\gamma,\mu) d\mu ]  \\&
		\hspace{2em}= \gamma\int_0^1\rho(\mu)p(\gamma,\mu)d\mu.
	\end{split}
\end{equation}
Based on the monotonicity and continuity properties of the cost function with respect to $\gamma$, we can derive that $\gamma^* =  \arg\min\gamma\int_0^1\rho(\mu)p(\gamma,\mu)d\mu$. As a result, the proof is completed.

\section*{Proof of Lemma 2}

The true reward probability associated with a given computation resources threshold is defined as $p$. This computation resources threshold has been selected for $n$ times, out of which $n_r$ instances have resulted in a reward. Therefore, its predicted reward probability is given by $\tilde{p} = \frac{n_r}{n}$. If $\delta$ can be determined such that $p\le \tilde{p} + \delta$, then $\tilde{p} + \delta$ is the upper confidence bound of the true reward probability $p$. According to Hoeffding’s inequality, $P\{ p-\tilde{p} \le \delta \} \ge 1-e^{-2n\delta^2}$. The inequality $p-\tilde{p} \le \delta$ always holds when $1-e^{-2n\delta^2} = \frac{N-1}{N}$, where  $\delta = \sqrt{\frac{\ln N}{2n}}$. As $n = N_t(\hat{\theta})$, $n_r = \sum_{n\in\mathcal{N}_t(\hat{\theta})}\mathbf{1}(\gamma_n \ge c(\theta_n,\mu_n))$. Then,  we have $\tilde{p} + \delta = \frac{\sum_{n\in\mathcal{N}_t(\hat{\theta})}\mathbf{1}(\gamma_n \ge c(\theta_n,\mu_n))}{{N_t(\hat{\theta})}} + \sqrt{\frac{\ln N}{2N_t(\hat{\theta})}}$. As a result, the proof is completed.

\section*{Proof of Theorem 2}

Let the optimal reward be $\gamma^*$, and the approximate optimal reward in the discrete space $R$ as $\gamma^+$, we have $\gamma^*+\xi \le \gamma^+ \le \gamma^*+2\xi$, where $\gamma^+\in R$. According to the definition of the minimum threshold for a client’s computation resources $\hat{\theta}$ with a fixed $\gamma$ and $\mu$, we can derive the following:
\begin{equation}
	\begin{cases}
		\hat{\theta}^*= \min \{\theta | \mathbf{1}(\gamma^*\ge c(\theta,\mu))\}), \\
		\hat{\theta}^+= \min \{\theta | \mathbf{1}(\gamma^+\ge c(\theta,\mu))\}). \\
	\end{cases}
\end{equation}
Let $(z-1)\xi\le \hat{\theta}^+ \le z\xi$, where $z\in\mathbb{Z}^+$, we have
\begin{equation}
	\begin{cases}
		(z-1)\xi = \min \{\theta | \mathbf{1}(\gamma^+ \ge c(\theta,\mu_r))\}), \\
		z\xi = \min \{\theta | \mathbf{1}(\gamma^+\ge c(\theta,\mu_l))\}), \\
	\end{cases}
\end{equation}
where $\mu_l < \mu < \mu_r$. According to the Bayesian Consistency and Bayesian plausible, the signals for $\mu_l$ and $\mu_r$ are given as follows:
\begin{equation}
	\begin{cases}
		\frac{
			\lambda(\underline{\tau})\rho(\mu_l|\underline{\tau})\underline{\tau} + \lambda(\bar{\tau})\rho(\mu_l|\bar{\tau})\bar{\tau}
		}{
			\lambda(\underline{\tau})\rho(\mu_l|\underline{\tau}) + \lambda(\bar{\tau})\rho(\mu_l|\bar{\tau})
		}
		=
		\mu_l,\\
		
		\frac{
			\lambda(\underline{\tau})\rho(\mu_r|\underline{\tau})\underline{\tau} + \lambda(\bar{\tau})\rho(\mu_r|\bar{\tau})\bar{\tau}
		}{
			\lambda(\underline{\tau})\rho(\mu_r|\underline{\tau}) + \lambda(\bar{\tau})\rho(\mu_r|\bar{\tau})
		}
		=
		\mu_r.
	\end{cases}
\end{equation}

Let $\rho(\mu_l) = \lambda(\underline{\tau}) \rho(\mu_l|\underline{\tau}) + \lambda(\bar{\tau})\rho(\mu_l|\bar{\tau})$ and $\rho(\mu_r) = \lambda(\underline{\tau}) \rho(\mu_r|\underline{\tau}) + \lambda(\bar{\tau})\rho(\mu_r|\bar{\tau})$, the approximate optimal signal is given by:
\begin{equation}
	\begin{cases}
		\rho(\mu_l|\underline{\tau})=\frac{\rho(\mu_l)(\overline{\tau}-\mu_l)}{\lambda(\underline{\tau})(\overline{\tau} - \underline{\tau})},\\
		\rho(\mu_l|\overline{\tau})=\frac{\rho(\mu_l)(\mu_l-\underline{\tau})}{\lambda(\overline{\tau})(\overline{\tau} - \underline{\tau})},\\
		\rho(\mu_r|\underline{\tau})=\frac{\rho(\mu_r)(\overline{\tau}-\mu_r)}{\lambda(\underline{\tau})(\overline{\tau} - \underline{\tau})},\\
		\rho(\mu_r|\overline{\tau})=\frac{\rho(\mu_r)(\mu_r-\underline{\tau})}{\lambda(\overline{\tau})(\overline{\tau} - \underline{\tau})},\\
	\end{cases}
\end{equation}
As a result, the proof is completed.

\section*{Proof of Theorem 3}

The optimal reward is denoted by $\gamma^*$, and the approximate reward is denoted by $\gamma^+$, where $\gamma^* +\xi \le \gamma^+ \le \gamma^* +2\xi$. Since higher reward leads to lower computation resources threshold in $\hat{\theta} = \min \{\theta | \mathbf{1}(\gamma\ge c(\theta,\mu))\}$ with a fixed $\rho$, it follows that $s(\gamma^*,\mu) \ge s(\gamma^+,\mu)$. Then, the difference in the server’s cost between the approximate optimal signal $\rho^+$ and reward $\gamma^+$, and optimal signal  $\rho^*$ and reward $\gamma^*$, is given by
\begin{equation}
	\begin{split}
		&\gamma^+ \int_0^1 \lambda(\tau)\int_0^1\rho^+(\mu|\tau) s(\gamma^+,\mu) d\mu d\tau
		\\& \hspace{2em} - \gamma^* \int_0^1 \lambda(\tau)\int_0^1\rho^*(\mu|\tau) s(\gamma^*,\mu) d\mu d\tau 
		\\& \hspace{2em} \le  (\gamma^* + 2\xi) \int_0^1 \lambda(\tau)\int_0^1\rho^+(\mu|\tau) s(\gamma^+,\mu) d\mu d\tau 
		\\& \hspace{2em} -  \gamma^* \int_0^1 \lambda(\tau)\int_0^1\rho^*(\mu|\tau) s(\gamma^*,\mu) d\mu d\tau 
		\\& \hspace{2em} \le \gamma^* \int_0^1 \lambda(\tau)\int_0^1\rho^+(\mu|\tau) s(\gamma^+,\mu) d\mu d\tau 
		\\& \hspace{2em} - \gamma^* \int_0^1 \lambda(\tau)\int_0^1\rho^*(\mu|\tau) s(\gamma^*,\mu) d\mu d\tau +2\xi 
		\\& \hspace{2em} \le 2\xi.
	\end{split}
\end{equation}
As a result, the proof is completed.

\bibliographystyle{named}
\bibliography{ijcai25}

\end{document}